\documentstyle[12pt,aasms4]{article}
\raggedright
\newcommand{\Delt}{$\rm \Delta$}

\newcommand{\MgL}{\rm $\rm M_{gas}/L_T$ }
\newcommand{\ML}{\rm $\rm M_T/L_T$ }
\newcommand{\MbM}{\rm $\rm M_{burst}/M_{gas}$ }
\newcommand{\MgM}{\rm $\rm M_{gas}/M_T$ }

\newcommand{\mss}{mag arcsec$^{-2}$}
\newcommand{\Bmoi}{\rm ${\mu_{B_{i}}}$(0) }

\newcommand{\lb}{$\langle \:$}
\newcommand{\gb}{$\rangle \:$}

\newcommand{\gt}{$> \:$}
\newcommand{\lta}{$\leq $}
\newcommand{\gta}{$\geq $}

\newcommand{\Bmo}{\rm ${\mu _B}$(0) }

\newcommand{\muo}{\rm ${\mu}$(0)}

\newcommand{\alp}{$\alpha$}
\newcommand{\etal}{{et al}\ }

\newcommand{\eg}{{\em e.\ g.\ }}

\newcommand{\Msol}{$M_{\odot}$}
\newcommand{\Zsol}{$Z_{\odot}$}
\newcommand{\Lsol}{$L_{\odot}$}

\input{epsf}
\input{rotate}
\input{psfig}
\begin{document}
\thispagestyle{empty}
\title{The Effects of Starburst Activity on \\
Low Surface Brightness Disk Galaxies}
\author{Karen O'Neil, G.D. Bothun, \& J. Schombert}
\author{Dept. of Physics, University of Oregon, Eugene OR, 97403}
\author{email:karen@moo.uoregon.edu}
\author{email:nuts@moo2.uoregon.edu}
\author{email:js@abyss.uoregon.edu}

\begin{abstract}
Although numerous simulations have been done to understand the effects of
intense bursts of star formation on high surface brightness galaxies, few
attempts have been made to understand how localized starbursts would
affect both the color and surface brightness of low surface brightness
(LSB) galaxies.   
To remedy this, we have run 53 simulations involving
bursts of star formation activity on LSB galaxies, varying both the
underlying galaxy properties and the parameters describing the
starbursts.  We discovered that although changing the total color of a
galaxy was fairly straightforward, it was virtually impossible to alter a
galaxy's central surface brightness and thereby remove it from the LSB
galaxy classification without placing a high (and fairly artificial) threshold for the 
underlying gas density. The primary effect of large amounts of induced
star formation was to produce a centralized core (bulge) component
which is generally not observed in LSB galaxies.  The noisy morphological
appearance of LSB galaxies as well as their noisy surface brightness
profiles can be reproduced by considering small bursts of star formation
that are localized within the disk.  The trigger mechanism for such
bursts is likely distant/weak tidal encounters.  The stability of
disk central surface brightness to these periods of star formation
argues that the large space density of LSB galaxies at z = 0 should
hold to substantially higher redshifts.
\end{abstract}

%%%%%%%%%%%%%%

\section{Introduction and Background}

Low surface brightness (LSB) galaxies have been systematically
under-represented in galaxy surveys due to selection effects whose
severity has not been properly appreciated in the past
(see reviews by Impey \& Bothun 1997, Bothun, Impey \&
McGaugh 1997).   Surveys to date (e.g. Bothun \etal 1986; Schombert etal 1990; 
Impey \etal 1996; O'Neil \etal 1997a) have identified three main classes of LSB 
galaxies: 1) dwarfs, defined by objects with scale lengths $\leq$ 1 kpc;
2) disk galaxies with scale lengths 1 $\leq$ $\alpha$ $\leq$ 5 kpc and
circular velocities in the range 80 - 200 km/s; 3) giant disk galaxies
with scale lengths $\geq$ 5 kpc.   (As documented by Sprayberry \etal
(1993), the properties of giant LSB disks are substantially different
from those of lower scale length (see also Knezek 1993, Pickering \etal
1997).)  In this contribution we are only concerned about LSB galaxies
that define the second category and which have central surface brightness
in the blue fainter than \Bmo \lta 23.0 \mss.  For these objects, 
multicolor photometry,
combined with 21-cm observations and H II region spectroscopy 
are consistent with their having an evolutionary path which branched
significantly from that which formed the traditional Hubble sequence (HSB) (i.e.
McGaugh \& Bothun 1994; de Blok, Bothun, \& van der Hulst 1995;
O'Neil \etal 1997b).  This suggests a perhaps fundamental
difference in star formation history between LSB and HSB disks.

To first order, however, the range of continuum colors between LSB and HSB disks
is essentially the same, which broadly means similar stellar populations.
The principle differences between the two systems are that 1) LSBs tend to
have higher fractional H I contents and 2) at a given circular velocity
($V_c$) LSBs have a lower stellar abundance (see data in McGaugh 1992;
McGaugh 1994).  At face value, this suggests that
LSBs are less evolved and have had less generations of massive star formation
than HSBs.  Prima facie evidence for this simple view comes from two
principle studies: 1) the measured  surface density of H I in LSB disks is
3-6 \Msol pc$^{-2}$ (Skillman \etal 1987; van der Hulst \etal 1993;
de Blok 1997; Pickering \etal 1997) -- well below the critical density for 
star formation (\eg Quirk 1972; Kennicutt 1989; Impey \& Bothun 1989);
2) a subsample of objects with velocities 3500-8000 km/s drawn from the
H II region spectroscopy of McGaugh (1992; 1994) have \lb log(O/H)\gb
=  $-$3.91 $\pm$ 0.30.  This sample consists of
26 individual H II regions in 12 host galaxies and the total observed
range  is $-$4.67 $<$ log (O/H) $<$ $-$3.55.  The host galaxies have  $V_c$
and dynamical masses comparable to L$_*$ HSB disks but an overall metal
abundance of Z $<$ 0.3\Zsol.

The surface brightness dependence of the luminosity-metallicity (L-Z) relation
for galaxies remains enigmatic.   When dwarf galaxies are included 
(e.g. Skillman 1998), it seems clear that luminosity and not surface
brightness is the principle driver.  However, Garnett \etal (1997) have shown
that for disk galaxies, the residuals from the L-Z relation do correlate with 
local disk surface density.   This suggests that while total mass 
(depth of the potential well) is the principle driver behind metal production,
it can be augmented by subsequent density-dependent star formation in the disk.
Our contention, based on the data, is that for disk galaxies with
scale lengths of 1--5 kpc, LSBs have systematically lower abundances
than HSBs.

It is possible this low metallicity is a result of a relatively young mean 
age for these systems as there has been insufficient time to produce many
metals.  In addition to being
consistent with their higher than average fractional gas content,
it also helps to explain the very blue colors of some disks.  These blue
colors are quite difficult to understand in terms of star formation alone,
given their low current star formation rates (SFRs) of
$\sim$ 0.1  \Msol yr$^{-1}$  (McGaugh 1992; de Blok 1997).  Moreover,
in $V-I$, LSB disks are often {\it bluer} than the most
metal-poor globular clusters, showing that the low metallicities alone cannot
account for the blue colors (McGaugh \& Bothun 1994).
Instead it
appears very blue LSB galaxies are among the least evolved objects known.  The best
example of this is provided by UGC 12695, a large, gas-rich LSB which
is perhaps the bluest disk in the nearby Universe (O'Neil \etal 1998a).

What we wish to explore in this paper is the coupling between LSB
and HSB disks.   Specifically, the measured SFR fails to produce the 
observed number of stars in LSB disks (e.g. their total luminosities)
by an order of magnitude.  This indicates that at times in the past
the SFR must have been substantially larger than its current value.
Given this, we probe the issue
as to whether or not episodic star formation in disk galaxies drives
sufficient excursions in surface brightness such that the typical disk
galaxy may go through alternating periods of being either LSB or HSB.
A priori, we know that this can not be the explanation for
the very blue disks.   However, a contingent of very red LSB galaxies has 
been recently discovered (O'Neil, \etal 1997a) which can be plausibly
identified with faded disks.  Between these extremes is a fairly continuous 
range of LSB galaxy colors, including an important  group whose V$-$I 
colors indicate an underlying old stellar population but whose U$-$B colors
indicate recent star formation in substantial excess of the average past
rate (O'Neil, \etal 1997a).  This mix of galaxy colors makes understanding 
the effects of starburst on LSB galaxies important to understanding  LSB
properties and morphology.

If LSB disks have experienced episodic star formation then this begs
an obvious question -- could LSB galaxies undergo significant bursts of
star formation and still retain their faint, diffuse appearance?  The aim of
this paper is to address that question by formulating a 2D model that
explores the effects of increased star formation activity at local places
in the LSB disk on its overall color and luminosity profile.  In
section two we describe our assumptions and underlying methodology.
Section three gives our computational results
showing how altering both the total mass undergoing starburst and the
underlying galaxy properties (\muo, $\alpha$, $\rm M_{gas}/L_T$, etc)
affect the final galaxy color and surface brightness.  Section four
discusses potential triggering mechanisms for LSB galaxy starbursts and compares
the results of our models with the observed LSB galaxy colors.

\section{Components of the Model}
\subsection{Physical Underpinning}

In addition to the differences cited in Section 1 between LSBs and HSBs, 
de Blok \& McGaugh (1997), based on dynamical data, strongly advocate that 
LSB disks have fundamentally lower surface mass densities than HSB disks 
(and LSBs may be more dark matter dominated).   This leads
to the physically reasonable situation that the density of gas follows the
density of stars ($(\delta \rho / \rho)_{gas} 
\propto (\delta \rho/ \rho)_{stars}$).
If the production of Giant Molecular
Clouds (GMCs) and subsequent massive star formation is density dependent,
this apparent physical difference between LSBs and HSBs may directly
translate into differences in star formation histories.  Certainly those
LSB systems measured to date seem to lie below the threshold column
density needed for the formation of GMCs.   

Studies of LSB disks have found them to be
deficient in molecular gas and dust (see Schombert \etal 1990;
Schombert \etal 1992; de Blok \& McGaugh 1997).  It should be 
noted, though, that it is certainly possible that small amounts
of $CO$ may have escaped detection due to beam dilution effects.  Indeed,
its these possible small scale regions of molecular material that may be
fueling the 2 - 4 individual H II regions usually observed in LSB disks.
The nature of these H II regions shows that stars of at least 50-70
\Msol are present so some massive star formation and metal enrichment
is occurring.  However, the filling factor of  H II regions is
very low currently and, given the SFR associated with these H II regions,
it is not possible that the entire stellar content of LSB disks can be
produced in this manner.  Hence, if large scale star formation in
any galactic disk requires a GMC component to the ISM then clearly
there must be times when an LSB disk has enough molecular material
to generate a burst of star formation. This argues that
$(\delta\rho/\rho)_{gas} \propto (\delta\rho/\rho)_{stars}$ is not
continuous but must bottom out at some {\it threshold}
$(\delta\rho/\rho)_{gas}$ so
that there remains an unused reservoir of sufficiently high column
density gas that can be converted to molecular material.

Limited observational support for this comes from various 21-cm
mapping of LSB systems (e.g. Skillman \etal 1987; van der Hulst 
\etal 1993; van Zee \etal 1997; Pickering \etal 1997) in which 
central column densities less than 10$^{20}$ cm$^{-2}$ are not
observed.  However, we emphasize that a representative sample of
LSB disks has not yet been mapped in H I and objects with lower
column density may yet appear (or be serendipitously detected in
the Parkes Multibeam survey - see Webster \etal 1998).  Nonetheless,  we adopt
this gas density threshold for a subset of our models
to determine the effects it may have on the starbursts.  These 
particular models, then, are
equivalent to allowing
any LSB disk in our model to experience elevated levels of star formation if
some agent acts to clump the gas regardless of \Bmo.

\subsection{Our Definition of a Starburst}

The purpose of our model is to explore the effects
that brief, intensive, but localized starbursts have on the color and surface
brightness profiles of typical LSB galaxies.   In particular, we wish
to determine if a starburst could, in fact, increase the
central surface brightness of an LSB disk and move it out of that domain.
At this point, its important
to clarify what we mean by ``starburst'' to avoid confusion later on.
Traditionally, starburst galaxies are morphologically distinguished by
strong central regions of star formation which typically increase the
bolometric luminosity of the host galaxy by a factor of 2-10.  Central
starbursts of this amplitude are usually triggered via galaxy interactions
or merging (see Mihos \& Bothun 1998; Smith \etal 1996).  Heckman \etal
(1998) have detected a very important attribute of these starburst galaxies,
namely, that it is only in very metal poor systems in which most of the flux
associated with the starburst escapes the galaxy at UV wavelengths.  
In metal-normal and metal-rich systems, most of the intrinsic UV flux
escapes the galaxy as re-processed Far Infrared radiation.  This result
strongly increases our expectation that starburst activity in a LSB 
disk would manifest itself mostly as an increase in the observed UV/blue
surface brightness.  Since LSB disks are generally observed to inhabit low 
density environments, the probability of their experiencing a strong
tidal encounter during a Hubble
time is low (i.e. Bothun \etal 1993).  Thus
we would not expect them to experience strong centrally
concentrated star formation and to become true starburst galaxies (see also Salzer 1998).
As a result, what we are exploring in this paper are the effects of non-centrally
located star formation bursts on the overall properties 
of an isolated, LSB system.
For the moment, we don't care how those bursts might be generated.

\subsection{The Model LSB Galaxy}

We have selected an initial LSB galaxy model based on the
mean characteristics of LSB galaxies as a class (O'Neil 1997; McGaugh 1992): 
1) no central bulge or bar, 2) low metallicity, 3) a surface density of gas
below the canonical threshold for star formation (see Kennicutt 1989).
Because the mass density in LSB disks is low (de Blok \& McGaugh 1997),
starburst activity can occur in a physical regime where the dynamical
time scale is longer than the duration of the starburst.   It should be noted that the physics
in this situation may be substantially different than when these timescales
are reversed in amplitude, such as in the case of ultra-luminous IRAS
galaxies (see Downes \& Solomon 1998; Mihos \& Bothun 1998).

For our models we represent the surface brightness distribution as an
exponential, e.g.

\[\rm \mu (r)\:=\:\mu(0)\:+\:1.086{r\over\alpha}\]

where \muo is the central surface brightness in \mss, $\alpha$ is the
galaxy scale length in kpc, and r is the radius in kpc.  The
central surface brightness and scale length were allowed
to vary between models, with \Bmo = 22.0 $-$ 24.5 \mss\ and \alp = 0.8 $-$
18.0 kpc, representative of LSB galaxies (i.e. O'Neil, \etal 1997b;
Pickering, \etal 1997; McGaugh \& Bothun 1994).  The outer edge of the
model galaxy, $r_T$, was set at the point where the surface brightness
profile drops to 27.0 B \mss.

LSB galaxies typically have a uniform color distribution throughout the
disk.   This is different than the case for HSB galaxies with similar
scale length where significant color gradients are often observed
(de Jong 1996).  Confusion over the color gradient issue stems from
the observations that giant LSB galaxies (those with scale lengths in
excess of 5 kpc) exhibit large color gradients with differences of
0.3 $-$ 0.5 mag difference in B $-$ R between the inner 2 scale lengths 
and the outer envelope (Bothun \etal 1990; Knezek 1993; 
de Blok, Bothun, \& van der Hulst 1995; Sprayberry \etal 1995).
But we are not modeling giant LSB galaxies here.  For the LSB disks
we therefore assume the underlying colors to be constant between
the inner and outer annuli of our defined localized starburst.  

The initial mass function (IMF) for each starburst in the model is of the
form:

$$\rm \phi(m)\:=\:m^{-x};\:\:\:x\:=\:\left\{ \begin{array}{ll}
0.25 &  m\: < \: M_{\odot} \\
1.35 &  M_{\odot}\: < \:m\: < \:2M_{\odot} \\
1.70 &  2M_{\odot}\: < \:m \end{array}\right.$$

where the slopes for the IMF were taken from Guiderdoni \&
Rocca-Volmerange (1987).  The IMF is normalized according to $\rm
\int_{m_l}^{m_u}\:m\phi (m)dm$ = 1, with $\rm m_{lower}$ = 0.55\Msol, and
$\rm m_{upper}$ = 120\Msol.  To account for the formation of brown dwarfs
and planets, 50\% of the star-forming mass is assumed to be converted into
non-H-burning objects (Bahcall, Hut, \& Tremaine 1985).  If this
assumption is incorrect, and more (or less) of the total mass is converted into
these objects, then the effects can be compensated for by varying the total
amount of the galaxy's mass undergoing starburst.  

\subsection{Star Formation in Model Cells}

For the purpose of forming stars, the model galaxy is divided into 30
slices in the radial direction and 30 in the angular direction for a total
of 900 cells.  The gas available for star formation in a particular cell
was determined from the observed range of total gas masses (or 
fractional gas content) found by de Blok (1997).
The underlying gas distribution was
assumed to follow the typical HI distribution of LSB galaxies --
flat out to 2$\alpha$, and then falling as $1/r$ (de Blok 1997, Figure
8.2).  The radial position of each cell is compared to the HI profile and a 
gas mass is assigned. 

Mihos, de Blok \& McGaugh (1997) have shown that a tidal
interaction between a LSB and HSB galaxy of similar mass does not result
in a central infall of gas, but instead may cause localized clumping of
the gas.  To model this type of behavior, we varied where the 
starburst center lies, letting it range between the true galaxy center and 85\% of the total
radius.  The strength of the starburst was usually assigned to be 15\% of the
galaxy's gas mass.  If the total gas mass in the cell containing the initial 
starburst was greater than, or equal to, the assigned starburst mass,
then the starburst was confined to that one cell.  Otherwise, the burst radius was
allowed to grow to encompass surrounding cells until enough gas mass had
been gathered to equal the total burst mass.   The starburst in the
surrounding cells was calculated such that the starburst mass
in a neighboring cell, $m(r)$, is given by $m(r)\:=\:{c \over r}\:*\:m_{initial}$
where, $m_{initial}$ is the mass in the initial cell of the starburst, $c$
is a constant which defines the spatial concentration of the burst 
and $r$ is the distance from the burst center.   

This scheme is a variation of that used by Mihos, Richardson, \& Bothun (1991) 
as, instead of tracking increases in density, we track increases in gas mass 
in each cell.  This is a concession to our ignorance of star formation in a low density environment and whether or not the Schmidt law would be applicable in 
an ISM in which the molecular gas appears to be either deficient or in very 
small scale clumps.
Phenomenologically then, since we clearly don't understand the physical nature
of the ISM in LSBs, it is easier to assess the effects of star formation 
activity  on LSB disk properties by just turning some cell gas mass
into stars.  This feature of our model, and how the properties of the
starburst relate to $c$ and $r$ parameters, 
is schematically shown in Figures~\ref{fig:concent} and \ref{fig:finalgas}.

As outlined in Figure~\ref{fig:concent},
the mass used to form stars in a particular cell will depend on how the
burst mass is distributed spatially within the galaxy.  Since the gas
density is low in LSB disks, often the highest stellar masses formed are
limited by the low probability of their being drawn from the small, finite gas mass
per cell.  
This is the unique feature of our model.
Increases in star formation activity in HSB disks are
generally thought to result from the build up of Giant Molecular Clouds
(GMCs) and there is sufficient mass to generate a relatively smooth,
continuous IMF.  However, the low gas densities in LSB disk force us to
to consider a situation
where the available gas mass is limited and thus the frequency of high
mass star formation may be inhibited by small number statistics.
Indeed  this effect may even
be the relevant physics (i.e. Oey \& Clarke 1998; Oey \& Kennicutt 1997).
The low gas density requires a significantly
larger scale length to reach the gas masses which are usually associated with
starburst activity in HSB disks.   In some cases, this scale length
is actually larger than the scale length of the underlying exponential
stellar distribution.  The physics of star formation in this regime
is currently under investigation (O'Neil, Carollo, \& Bothun 1998).

This starved gas mass per cell has the physical effect that our models
have difficulty in creating stellar masses \gta 2.2\Msol.  
Consequently, without a well populated upper main sequence per 
star formation event, there can be little chemical enrichment and
little alteration by the starburst of the pre-existing stellar
population's color or surface brightness.   
This situation is not encountered if we use the threshold approach
that fixes a minimum value of  $(\delta\rho/\rho)_{gas}$.  In
this case, the gas mass per cell which is available will not have
as strong of surface brightness dependence and subsequent star
formation events will have stronger effects on the overall properties
of the LSB disk.  We will run both sets of models
and discuss which of them are most consistent with the
observed properties of LSB disks.

Once the appropriate amount of mass in the cell has been
converted into stars using the input efficiency, the total luminosity and
color are determined from the IMF.  This population is then allowed to
evolve using the standard stellar evolution prescriptions.   
The low metallicity of LSB disks constrains our
choice of stellar evolutionary tracks to those with stars having Z=0.001
and Y=0.30 (obtained from Becker 1981; Mengel, \etal 1979; Becker \& Iben
1979; Schaller, \etal 1992; and Sweigart \& Gross 1978).  Once formed in
the model, stars were allowed to evolve from the main sequence through
their carbon-ignition phase.  The original galaxy colors are an input to
the model and no color evolution was assumed for this underlying stellar
population.  Over the duration of the starburst, the evolution of this
underlying population is negligible.

As a check that our basic modeling and stellar evolutionary code procedure 
is valid, we simulated a true starburst by assuming all the gas was in 
``one" cell and computed, in effect, a global change in color and luminosity 
as a function of percentage of gas mass involved in the burst.  These results 
are shown in Table~\ref{tab:paramch} and are completely
consistent with standard starburst models (i.e. Fritze-V. Alvensleben
1998).  When applied to the low
density ISM of LSB disks, we spread these conditions over many cells.
Depending on the choice of model parameters, this produces a range of
probabilities  of massive star formation.  For most choices, this probability
is small, unless the gas density threshold approach is used.

\section{Model Results}
\subsection{Overview}

The primary focus of our models is to understand how an occurrence of
star formation activity in some local region of a LSB galaxy affects
it's overall color and surface brightness profiles.  Since 
LSB galaxies vary greatly in size, luminosity and density, knowing the
effects of altering many of the galaxies' other parameters (M/L, $\rm
M_{gas}/M_T$, \alp, etc) is an important step to understanding the
starburst process.  To this end, a series of models were run altering only
one variable at a time.  The actual model parameters are given in
Table~\ref{tab:model}, and a summary of the results are given in
Table~\ref{tab:param}.  Additionally, to give an idea of the effect the starbursts
have on the gas density of the galaxy, a plot showing the before starburst
and after gas distributions are shown in Figure~\ref{fig:finalgas} for
a number of models.

All models were run with t$\rm _{starburst}$ = 2 Myrs and \Delt t =  1 Myrs.
After 100 Myrs, the stellar population produced by the burst has faded
and blended into the underlying stellar population.  (See
Figure~\ref{fig:fig1} which shows the changes in $U-B$ color with time.)
These colors are always
expressed in contrast to the original galaxy color, also a variable in
Table~\ref{tab:model}, and are plotted as $\Delta(U-B)$ as a function of
time.  The blue surface brightness profiles and radial color profiles at a
timestep of 2 Myrs are shown for every model in Figure~\ref{fig:fig1}.
The colors are plotted relative to the initial galaxy color,
$\Delta(U-B)$ as a function of radius.  The individual model results
plotted in each of these figures are discussed below.

\subsection{Central Surface Brightness}

We begin by considering three LSB galaxies with initial central
surface brightness \Bmoi = 23.0, 24.0 and  25.0 \mss\ and we run them
through models with and without the HI column threshold.
Under the no threshold model, the smallest color change (only
$-$0.04 in $U-B$) occurred in the \Bmoi = 25.0 \mss\ model.
Recall, that the no threshold model explicitly assumes
$(\delta\rho/\rho)_{gas} \propto (\delta\rho/\rho)_{stars}$.
In this case, the low assumed gas density plays a very large role.
Our procedure places gas mass in each cell, with the amount dependent on the 
input $\rm M_{gas}/M_T$ and the HI profile.  Since $M_T$ is smallest
in the \Bmoi = 25.0 \mss\ model (because there are less stars and \alp\ 
is constant), there is less gas in each cell for a fixed $\rm M_{gas}/M_T$.
This substantially reduces the probability for massive star formation.
On the other hand, if we use the threshold model, setting the gas density
of each model to the density obtained for a \Bmo = 22.0 \mss\ model,
we produce a result
that agrees with our intuition.  That is, the burst of star formation
does produce the most dramatic changes in contrast to a surrounding lower
surface brightness stellar population, resulting in the \Bmoi=25.0 \mss\ model
having the largest color change.   These two situations are summarized in
Table~\ref{tab:modell} where the a1-a3 models are for the no threshold case and
models l1-l4 have the threshold.

To simulate the effects of increasing gas mass, we ran models with
and without the gas density threshold but varied the fractional gas content
$\rm M_{gas}/M_T$ or the baryonic mass fraction $\rm M_L/M_T$. 
Not surprisingly, rising gas masses led to a higher frequency of high mass 
star formation and bluer overall colors.   For instance,
for \ML = 10.0 \Msol/\Lsol, a typical value
for LSB galaxies (see de Blok \& McGaugh 1997), the change in U$-$B is
-0.20, and could easily be increased to -0.34 with a small change in \MgM
from 0.05 to 0.10. 
We also ran an extreme case of the gas density threshold model in which 
75\% of the available gas was allowed to undergo a starburst.
This model was run with \Bmoi = 24.5 \mss and the spatial concentration (c) 
was set to 50 arcsecond..
The result is the galaxy forms a r$^{1/4}$ (bulge)-type profile.
That is, the galaxy behaved the same as a galaxy which collapses or otherwise
undergoes a considerable infall of gas to its core.

Establishing the H I column density threshold for models l1-l4 carries
with it a hidden assumption namely that the galaxy has a significant
reservoir of gas, spread throughout its disks, which is available for
starburst activity that will elevate the overall surface brightness
level.  Why is it then that so many
of the very bluest LSB galaxies can have \Bmo  = 24.0 -- 25.0 \mss?  If
they have, in fact, been brightened significantly then this implies
there is a progenitor population of extremely LSB disks (e.g. 
\Bmo $<$ 25.0 \mss) that have large amounts of gas.  Its likely
that such a population would have been detected in blind 21-cm 
surveys by now (e.g. Briggs 1990; Spitzak \& Schneider 1998).  
Furthermore, the red (e.g. $B-V$ $>$ 1.1) LSB disks discovered by
O'Neil \etal (1997a) inhabit the same range of \Bmo as the blue LSB
disks.  Overall, the lack of any \Bmo vs. color relation makes
any fading scenario unlikely and the gas density threshold scenario 
problematical.  Models without the gas density threshold  would tend to
preserve this observed non-correlation.

Now we emphasize that, to date, no  \Bmoi \lta 23.5 \mss\ {\it disk} galaxy
has yet been mapped in H I so we really can't assess the validity of
the gas density threshold model.  As a result, the majority of the subsequent models 
were run with no threshold criteria.
If our assumption is wrong, and LSB galaxies do indeed have a minimum
gas density which is independent of the actual value of \Bmo, 
the results can be readily determined by simply adding the
appropriate l model results to the model in question.   We assess
the likelihood of the minimum gas density model in Section 4 as the
results in Figure~\ref{fig:fig1} clearly indicate that it does lead to substantial
changes in \Bmo.

\subsection{Scale Length}

Starting now with  \Bmoi = 24.5 \mss\ and no gas density threshold criteria, models
b1 - b4 varied only in the initial 
value of the scale length, \alp, from 0.8 kpc to 10.0 kpc. 
The sharpest color changes with time are
for the galaxies with the largest scale length, which is well matched by the limited observational 
data (see de Blok \etal 1995) which shows that large scale length LSB disks tend
to have relatively strong color gradients.
Moreover, the \alp\ = 10 kpc model shows a strong blue core.
This behavior is highly reminiscent of the class of giant low
surface brightness galaxies described in Sprayberry \etal (1995), all
of which seem to have a pronounced bulge component (like Malin 1 - see
Impey \& Bothun 1989) which could be the faded remnant of this blue core.

Changing the scale length is equivalent to altering the total luminosity
and mass of the galaxy.  However, given our prescription of how the
size of the starburst is determined by the need to gather enough mass to
reach the variable $\rm m_{burst}/m_{gas}$ (usually 15\%), the behavior of
the starburst population will vary with the mass within the cells.  When
the galaxy is large (large \alp), then the mass in the cells is high and
the starburst is localized. Thus, sufficient numbers of high mass stars
are formed per cell, which leads to a bluer integrated color.  However,
when the galaxy is small, the starburst is spread over a large area and
the mass per cell is small and the frequency of forming high mass stars
(above 2.2\Msol) is decreased.
The radial color profile also exhibits the effect of a localized versus
widespread starburst.  The small scale length models have radial color
profiles that show the starburst remained near its point of origin (always
2\alp for these models).  The profiles display small blue haloes with red
cores.   This is reminiscent of the color profiles seen in some
nucleated dwarf galaxies in Virgo and Fornax (e.g. Caldwell \&
Bothun 1987; Ichikawa \etal 1986; Han \etal 1998). 

\subsection{Starburst Location and Concentration}

In order to understand the effects of the specific location of the
starburst on LSB
galaxies, we ran a series of simulations, models c1 - c5, varying the
radius at which the starburst occurred, $\rm r_{burst}$.  In model c1, the
simulation was run with 15\% of the mass in each cell undergoing
starburst, resulting in an even star formation distribution.  In models c2
- c5 we let $\rm r_{burst}/r_T$ vary as 0.10, 0.30, 0.50, and 0.80,
respectively. 

The interpretation of these models is fairly straight-forward, with the
starburst activity being very centralized for $\rm r_{burst}/r_T$ \lta
0.50. The gas density is high in the core, which produces some high
mass star formation at that location.  The result is a bluer, higher
surface brightness core component to our model LSB disk.
The overall change in color
(\Delt (U$-$B)) was significant, ranging from
$-$0.27 to $-$0.19.  The change in the surface brightness profile was also
dramatic, with increases of 0.5 to 1.0 B \mss\ in the inner
regions.  This would have a noticeable impact on the morphological
appearance of the galaxy via the rapid development of a bulge or central
bar.  In general, these central structures are not observed in most
LSB disks implying that this mode of star formation does not occur in
these systems.
As $\rm r_{burst}/r_T$ increased above 0.50, the starburst activity became
non-centralized, producing small bumps in the color/surface brightness
profiles with amplitudes of $-$0.08 in $U-B$.  The change in their surface
brightness profiles was also minimal. Thus, we conclude that the burst
location is critical to the evolution of a LSB disk.  

The next parameter we varied in our models was the amplitude of the
concentration parameter, $c$.  
Because this parameter is artificial, representing the
unknown and highly varying triggering force, we allowed $c$ to range from
1.0'' to 100.0''. (Recall that c, like the radius, is kept in arcsec for
the purpose of modeling.  See section 2.2 for more information.)
 When $c$ was at its lowest value (c=1.0'', model f1), the
starburst spread across the entire galaxy, and thus inevitably concentrated at
the galaxy's core resulting in a sharp change in the galaxy's color
profile and a measurable change in the galaxy's inner surface brightness
profile (see Figure \ref{fig:fig1}).   As expected,
this scenario comes closest to
the global starburst scenario which can indeed transform a LSB disk into a
disk of significantly higher surface brightness, and has possible
application to the detection of faint blue galaxies at intermediate
redshift.  We hasten to add, however, that most LSB galaxies at $z=0$ are
not in an environment which is conducive to bursts which
drive $c$=1.0 (see Bothun \etal
1993; Mo, McGaugh, \& Bothun 1994).  With $c$ \gta 5.0'', the color profiles show only a
localized change at the radius of the initial starburst and the effect
on integrated color is substantially reduced (by 0.20- 0.30 mag in $U-B$).
No measurable differences in \Bmo could be detected for the c \gta 5.0'' models.

\subsection{Initial Stellar Population}

In an attempt to understand how starbursts would affect galaxies at
different stages in their evolution, we altered the colors of the initial
stellar population, from very blue ($U-B = -0.20$, $B-V = 0.40$, $V-I =
0.80$) to the very red ($U-B = 0.60$, $B-V = 1.50$, $V-I = 2.00$).  The
results of this simulation are shown in Figure~\ref{fig:fig1} (models
h1-h5).  The total change in color between the five models is significant,
with the largest change always occurring in the U$-$B color, and the
smallest in V$-$I.  Because the effects of starburst are less significant on
galaxies which are already fairly blue, we chose colors slightly redder
than the average LSB galaxy color for most of our simulations (i.e.  U$-$B =
0.2, B$-$V = 0.8, V$-$I = 1.2) to match the colors of new LSB galaxies currently
being discovered in CCD surveys (O'Neil \etal 1997a).

The starbursts are more easily detected in the radial color profiles for
reddest underlying stellar populations.  This is simply due to the enhanced
contrast bright, blue stars have against an older mix of stars.
Additionally, it should be noted that the color change in the red models
is far larger in U$-$B than in V$-$I, consistent with notion the  $V-I$
is a reasonable metallicity indicator due to its lower sensitivity to
newly formed stars.  Colors that are consistent with the model $U-B$ and
$V-I$ values  have been detected by
O'Neil \etal (1997a; i.e. C1-1 with U$-$B = 0.13,
and V$-$I=1.20).

\subsection{A Flat IMF}

In the no gas density threshold models, the gas density of the system is
too low for the formation of many massive stars.  We can force this situation
to change by considering a flat IMF.  This was done using a series of models
which put m$\rm _{lower}$ = 1.1\Msol in the IMF.  These five models were run to
mimic the possibilities (and extreme cases) of all the above models.
Models j1, j2, and j3 have localized starbursts occurring at r = 0.60r$_T$
(j1 \& j2) and r= 0.10r$_T$ (j3), while models j4 and j5 have starbursts
spread evenly throughout the galaxy.

Models j1 \& j4 used the average values for all of the models, while
models j2, j3, \& j5 used an exceptionally high M$_{gas}$/M$_T$
(M$_{gas}$/M$_T$ = 0.15).  Figure~\ref{fig:fig1} a shows the surface
brightness profiles of these models after 2 time steps.  What can
clearly be seen is that forcing the galaxies to produce high mass stars
raises the surface brightness of the burst regions.  In all cases, the
central surface brightness of these galaxies is raised by at least 0.1
\mss, and is raised by 0.4 \mss\ for model j5.  Given the observations that
most LSB disks (e.g. those with central surface brightnesses fainter
than 23.0 \mss) are already quite blue, this scenario can't really
happen because it would represent the case of an LSB transitioning to
a higher surface brightness disk which would have predicted colors
much bluer than are generally observed (e.g. galaxies like M101 don't
have $B-V$ \lta 0.3).  This, coupled with the observed low
metallicity and dust content of these systems, strongly suggests that flat IMF
star formation in a metal-poor environment is unlikely to be occurring.

Figure~\ref{fig:fig1} shows the color changes of these galaxies
with time.  As expected, forcing the galaxies to produce high mass stars
also causes significant fluctuations in the galaxy colors with time.
Again, these changes would be readily detectable, but considering the
large spread in colors (up to $-$1.33 in $V-I$) it may be difficult to
determine the true state of a galaxy in this stage.
As a last, extreme test of the effects of starbursts on LSB galaxies, we
returned m$_{lower}$ to 0.55\Msol and 
let the c parameter go to an extremely high value (c=50.0'') which, in
effect, allows as much mass in each cell as was available to undergo
starburst. At the same time, \MbM was allowed to vary from 0.25 (model i1)
to 0.75 (model i3), and the total \MgM was varied from 5\% (models i1-i3)
through 50\% (model i5).
The change in the surface brightness profiles for these models
was significant, as was the total change in color (\Delt (U$-$B) = -0.28 $-$ -0.86;
\Delt (B$-$V) = -0.18 $-$ -0.88; \Delt (V$-$I) = -0.10 $-$ -0.94; for
models i1 -- i5, respectively).  It should be noted
that these models are unrealistic cases but do serve as valid boundary condition
checks of our model procedure. That is, if we tell the model to simulate
a global starburst, then it does.

\section{Star Formation in LSB Galaxies}
\subsection{\bf Can {\bf \Bmo\ {\bf Significantly Change?}}}

The motivation for our models is to assess the degree to which disk
surface brightness is stable to star formation activity.  This issue
is important in understanding the space density of galaxies as 
a function of surface brightness and what the overall evolution of
this bivariate luminosity function with redshift might be.  We 
have run two general sets of models, one in which the surface density
of gas steadily decreases with decreasing \Bmoi and one in which
a gas density threshold surface density is established which remains constant
as \Bmoi continues to decrease.  The no gas density threshold models generally
have sufficiently low surface gas densities that the formation of
stars with mass \gta 2.2\Msol is inhibited in 
most of the cells where we have introduced a perturbation
(see Figure~\ref{fig:stars}).  In this case, although the structural nature
of the changed due to the starburst activity (for example, gaining a centralized
bulge or core), {\it the measured central disk surface brightness for these
galaxies never changed significantly}.
To verify this result, we re-ran all of the models with \MgM = 10\%
and \MbM = 20\%.  The results were the same -- although allowing larger
percentages of mass undergo starburst increased the total change in
color in each model, there was still no change in the central
surface brightness of the galaxies.  In fact, once \Bmo has been specified for a
galaxy, altering that value in any significant way was difficult if not
impossible to obtain.

The models with no gas threshold show the robustness of \Bmo
when star formation is occurring in our model low density environment.
To put it another way, whenever we force an intense starburst either
by a) raising the gas mass to high levels, b) forcing the burst to
be centralized either by varying $\rm r_{burst}/r_T$ or $c$ or c)
forcing a flat IMF to overcome the cell tendency of not
producing high mass stars, we generally produce another structural
component to the galaxy.  The observation that LSB disks, in general, have
no bulges, bars or nuclear activity would strongly argue that such a
centralized starburst never occurred in these systems.  The robustness of
\Bmo to localized star formation events and/or smaller global events is
consistent with the high space density of LSB disks now observed (e.g.
McGaugh \etal 1995; Sprayberry \etal 1997a, 1997b; Dalcanton \etal 1997). 

As
surface brightness is determined by the convolution of the mean luminosity
of the stellar population and the average separation between stars, our
model results would seem to indicate that changes in mean luminosity are
more than compensated for by whatever structural conditions existed during
the formation epoch of these disks that allowed for relatively low surface
number density of stars (see O'Neil 1998).  That is, once the basic
structure has been laid down, it would seem that increasing \Bmo through
modest star formation events is extremely unlikely.  Thus,
there should be little evolution in the distribution of \Bmo\ as a function
of redshift out to modest redshifts (z $\sim$ 1).  This implies that
the large space density of LSB disks at z = 0 should be preserved out
to z $\sim$ 1.  The Sloan Digital Sky Survey should ultimately produce
a data set that either confirms this expectation or demonstrates that
our assumptions about the nature of starburst in LSB disks are wrong.

One of those incorrect assumptions may be the gas density in the
no threshold models is simply too low to produce much
luminosity per star formation event due to a dearth of massive star
formation.  Thus the only means of obtaining an increase in the galaxy's 
disk central surface brightness is to allow a significant  increase in 
gas density within the underlying galaxy.  Within our models this 
can be achieved through decreasing the number of cells within the galaxy 
and thereby increasing the total mass within each cell.  This is equivalent
to establishing a threshold column density within 2-3 scale lengths
independent of \Bmo.

In models k1 -- k4 the galaxy was broken up into only 25 cells, thereby allowing
for considerably higher concentrations of mass than was available in all the
previous models.  The subsequent starburst had bursts strength akin to that
in the i models, with \MgM = 5\% -- 50\%, c = 50.0'', and \MbM = 25\% -- 50\%.
Additionally, the starburst was spread evenly throughout the galaxy,
in the same manner as model c1.  The extreme burst strength was necessary 
to successfully increase \Bmo by a measurable amount, while the even
distribution of the starburst was done because of the low resolution of the 
model -- since the galaxy was broken into only 5 radial pieces the distinction 
between having the starburst occur at 0.60R$_T$ and 0.30R$_T$ is minimal.

The resultant surface brightness profiles of models k1 -- k4 are given in 
Figure~\ref{fig:fig1}.  What should be immediately noted is that even in the 
case of model k1, where 25\% of the gas of these artificially condensed galaxies
was allowed to undergo starburst, the central surface brightness increased by only
0.5 \mss. In order to obtain an increase of 1 \mss\ in the central disk surface
brightness, 50\% of the galaxy's total gas content had to undergo a starburst.
Thus, in order to achieve a measurable increase in the central surface 
brightness, either inordinate amounts of gas must be added to the galaxy's core
or a substantial pre-existing reservoir of column density $\sim$ 10$^{20}$
must exist independent of \Bmo.

\subsection{Triggers for Gas Flows}

Is it reasonable to expect a delivery mechanism that could
introduce large amounts of gas to the core regions?  Internal mechanisms
such as bar driven flows seem quite unlikely as LSB disks generally
don't have any bars.  Stochastic processes, such as those operating
in irregular galaxies, probably also do not work since LSB disks
are strongly rotationally dominated.   Thus the only real possibility  is
tidal interactions.  The connection between
interacting galaxies and SFRs has been well documented both
observationally and through computer simulations.  Larson \& Tinsley
(1978) showed that the very blue colors of some galaxies could be
explained if those galaxies were undergoing tidally triggered bursts of
star formation.   This mechanism was shown to effectively work in
spiral-rich clusters of galaxies (e.g. Bothun \& Schommer 1982) where
many HSB spirals are located.  More recent studies have
shown that gravatationally interacting galaxies have a higher (average)
SFR than their more isolated counterparts (i.e.  Kennicutt, \etal 1987;
Bushouse 1986).  Additionally, Mihos, \etal (1991) used simulations to
show that, with high surface brightness (HSB) galaxies, tidal interactions
between nearby (distance = 12 --36 kpc) non-merging galaxies can result in
starbursts involving 12\% -- 24\% of the galaxies' mass (see also Mihos
\etal 1993; Mihos \& Hernquist 1994, 1996).

External triggers for star formation in LSB galaxies, though, are most
likely provided by relatively distant tidal disturbances.  HSB disks are
generally located in an environment which is favorable for a few strong
tidal encounters over a Hubble time.  These encounters typically cause a
central inflow of gas and often the formation of an inner bar.  Increase
in star formation in HSB galaxies from tidal interactions is therefore
centrally concentrated and an appreciable change occurs at the core of the
interacting HSB galaxies. LSB galaxies, though, have been shown to be
stable against the growth of bar formation and large scale central inflow
of gas during tidal encounters but not necessarily stable against local
instabilities.  Mihos, de Blok, \& McGaugh (1997) simulated a collision
between a HSB and LSB galaxy with similar properties. Soon after the
closest approach, the HSB galaxy developed a strong bar which persisted
through the end of the simulation and which presumably triggered a central
inflow of gas and a strong nuclear starburst.  Although being strongly
perturbed, the LSB galaxy did not form a bar and therefore would not have
undergone a similar central starburst.  Instead the encounter excited star
formation throughout the disk or in locally concentrated regions away from
the galaxy's center.  It is probable, then, that the more distant tidal
encounters LSB galaxies experience result in local, non-centralized
instabilities in LSB galaxies, and therefore in the triggering of
non-centralized starbursts which we modeled in the previous section.

An important note for this discussion is V\'{a}zquez and Scalos's (1989)
finding that starbursts do not typically occur during the gas compression
stage but in fact occur well after the gas has re-established.  In other
words, the V\'{a}zquez and Scalo model suggests that some disks can
have tidally induced star formation well after perihelion.  Thus
although LSB galaxies are often more isolated than their HSB counterparts
(Bothun, \etal 1993), 
this does not preclude the galaxy from being in the midst of
tidally induced starbursts.  V\'{a}zquez and Scalo show starbursts occur
20$\tau_C$ -- 80 $\tau_C$ after the initial gas density increase.  Here
$\tau_C$ is the typical collision timescale which is likely quite long
(\gt 10$^8$ yr) for the diffuse LSB galaxies being studied.  As many of
our detected LSB galaxies are located in the outskirts of spiral rich
clusters, the V\'{a}zquez and Scalo scenario may well apply.

Identifying galaxies which are currently undergoing or have recently undergone
star formation is difficult.  As is readily seen in Figure~\ref{fig:fig1},
stochastic star formation could easily produce structural
noise and hence one possible manifestation of this process is
the fairly noisy nature of the surface brightness profile and/or the
optical appearance of the galaxy.  Good examples of this are shown in
Figure~\ref{fig:gal} (from O'Neil, \etal 1997a, 1997b).  P2-4, with \Bmo =
25.1 \mss\ and its highly clumped morphology, is an extreme example of
this.  It completely lacks any centralized core, and consists primarily of
localized regions of above average stellar density.  We would identify
these as local star forming clumps (the clumps are rather blue).  P3-3 is
a more common example of a LSB galaxy, with has a definite exponential
surface brightness profile and fairly even color profile, but with
numerous bumps and wiggles present that are likely manifest of an
irregular distribution of recent star formation.  

For the red LSB galaxies we definitely detect a group with B$-$V
and V$-$I colors indicative of an old stellar population but U$-$B colors
indicating current star formation (see O'Neil \etal 1997a).
If these galaxies had colors similar to those of
very red LSB galaxies and then underwent a burst of star formation, their
colors could easily be changed by up to \Delt (U$-$B) = $-$0.4 (for a high
\MgL ratio), while their V$-$I and \Bmo are virtually unchanged.
An example of this can be seen in models h4 and h5, where the final galaxy 
colors are fairly close to the colors of U1-8 in O'Neil \etal 1997a
($U-B$ =0.31, $B-I$ =2.79), with final colors of 
U$-$B=0.04, B$-$I=2.3 and U$-$B=0.26, B$-$I=3.3 for models h4 and h5, respectively.

These changes easily explain the colors of U1-8 and similar galaxies, and
can account for part of the much low U$-$B colors of C5-3.  It is thus
likely these galaxies are older galaxies which are now or have recently
experienced a tidal interaction which triggered significant starburst.
Most of these galaxies are located in the outskirts of the Pegasus and
Cancer clusters of galaxies.  Weak/distant tidal encounters would be
expected in such an environment.

\section{Conclusion}

\hskip 0.2in We ran 53 different simulations, varying all possible galaxy
parameters and allowed up to 75\%  of the galaxy's gas to undergo
starburst.  While we could significantly change the galaxies' total color,
creating very blue LSB galaxies,
it was virtually impossible (without placing a high threshold criteria
for the gas) to significantly alter the galaxy's central
surface brightness.   Instead the primary effect of large amounts of induced
star formation was to produce a centralized core (bulge) component.
While some LSBs have this component (e.g. Malin 1), most are devoid
of any central luminosity excess above the fitted exponential which
strongly suggests that such centralized bursts didn't happen in these
systems.  We also suggest  that LSB galaxies
evolve through sporadic bursts of star formation and that the colors and
noisy morphologies displayed by many detected LSB galaxies can be
explained by recent starbursts triggered through distant/weak
tidal interaction.   Since the observed current SFRs are an order of
magnitude too low to produce the observed luminosity in LSB disks, it
seems clear that some sort of episodic star formation has occurred.

Our modeling procedure has assumed that $(\delta\rho/\rho)_{gas} \propto (\delta\rho/\rho)_{stars}$ in LSB galaxies.  De Blok \& McGaugh (1997) have
already shown, through analysis of rotation curves, that the low surface
light density of these systems does translate into low surface mass
density.   We believe this to be the basic physical difference between
HSB and LSB disks that should directly translate into different star formation 
histories.  Of course, one should question this assumption as it
potentially leads to the following dilemma:
if ($\delta\rho/\rho)_{gas} \propto (\delta\rho/\rho)_{stars}$
in a continuous manner then how come do the very low \Bmo systems have
any stars in them at all, provided a molecular cloud medium is a prerequisite
for large scale star formation to occur in {\it any} disk galaxy.?
On the other hand
if $(\delta\rho/\rho)_{gas}$ is the similar between
high and low surface brightness disks, one is very hard pressed to
understand why star formation appears to be so different in the LSBs
and/or the lack of dust/heavy elements in LSBs relative to HSBs of
the same $V_c$.   

We suspect that the actual truth lies somewhere between these two
extremes but that truth will be elusive.
While this is the subject of a larger investigation (O'Neil, Bothun
\& Carollo 1998) the executive summary is that, in a LSB disk, to
obtain the equivalent gas mass which is, say, contained in a GMC in
our Galaxy, requires a significantly larger length scale.   When the
disk of our model LSB galaxy is broken up into cells, the net result
of this large length scale requirement is that the probability of 
massive star formation per cell is low.  This predicted dearth of massive
star formation, of course, is consistent with the low metallicity and
dust content observed in LSB disks and would result in a very slowly
evolving population but leaves open the question of how such low
surface density disks could have formed in the first place.

To overcome this dearth of massive star formation we have run a set
of models that fixes a threshold column density of H I that remains
so even as \Bmoi decreases.  In this case, there is significantly
more gas available for star formation and some of the limitations of
the previous models are overcome.   However, in most cases the
threshold models cause an inner r$^{1/4}$ component to develop.  Thus
the LSB disk gains a ``bulge'' in response to the starburst and this
is generally not observed.  To effect a large increase in \Bmo with
a localized disk starburst event generally requires an inordinate amount of
gas to be converted into stars.  Moreover, its unclear that the 
threshold model is able to preserve the observed and important non-correlation
between \Bmo and disk color or account for the offset in mean log O/H
at a given $V_c$ with respect to HSB galaxies.  Most worrisome about
the threshold model is its implication that LSB disks of arbitrarily
low \Bmo are sitting around with ample amounts of gas.  To date,
systems like this have not been discovered in blind H I surveys.
Hence we believe that the slow evolutionary rate of LSB disks is likely
controlled by low gas density that accompanies low \Bmo which precludes
the formation of very many massive stars per star formation event.

Our primary result from the modeling procedure is that once \Bmo\ is
established for LSB systems, it is extremely difficult to alter it.
That is, disk systems are quite unlikely to hop back and forth between
states of high and low surface brightness due to episodic star formation.
We thus conclude that if a galaxy forms as a LSB
galaxy, due to low gas density, environmental conditions, etc, it will
remain a LSB galaxy barring any major encounter catastrophe.  This
implies that the large space density of LSB galaxies at z = 0 should
hold to substantially higher redshifts.  This may have relevance to
understanding the nature of QSO absorption line systems at these
redshifts (see Linder 1998).  This also suggests that deep CCD
surveys should reveal this population, if those surveys are relatively
free of selection effects.  This selection effects, of course, will
be more severe than those associated with finding z=0 LSB systems
due to the significant (1+z)$^4$ dilution factor.

\clearpage
\centerline{References}
Alvensleben, F-V. 1998, preprint\\
Bahcall, J., Hut, P., \& Tremaine, S. 1985, ApJ, 290, 15\\
Becker, S. 1981, ApJS, 45, 475\\
Becker, S. \& Iben, I. 1979, ApJ, 232, 831\\
Bothun, G.D. \& McGaugh, S 1998 in preparation\\
Bothun, G.D., Impey, C., \& McGaugh, S. 1997 PASP 109, 745\\
Bothun, Gregory D., Schombert, James M., Impey, Christopher D., Sprayberry, David, \& 
McGaugh, Stacy S.  1993 AJ 106, 530\\
Bothun, G.D., \etal 1990 ApJ, 360, 427\\
Bothun, Gregory D., Mould, Jeremy R., Caldwell, Nelson,
\& MacGillivray, Harvey T.  1986 AJ 92, 1007\\
Bothun, G.D., \& Schommer, R.A. 1982 AJ 87, 1368\\
Briggs, F.H. 1990 AJ, 100, 999\\
Bushouse, H. 1986, AJ, 91, 255\\
Caldwell, N., \& Bothun, G.D. 1987 AJ 94, 1126\\
Davies, R. L., Bertschinger, E., \& Baggley, G.  1993 MNRAS 262, 475\\
Dalcanton, Julianne J., Spergel, David N., Gunn, James E., Schmidt, Maarten,
\& Schneider, Donald P.  1997 AJ 114, 63\\
De Blok, W. 1997, Ph.D. thesis University of Groningen, Groningen, Netherlands\\
De Blok, W., \& McGaugh, S. 1997, MNRAS 290, 533\\
De Blok, W., Bothun, G., \& van der Hulst, J. 1995,  MNRAS, 274, 23\\
De Jong, R.S. 1996 A\&A, 313, 377\\
Downes, \& Solomon, 1998 ApJ, in press\\
Garnett, D. R., \etal 1997 ApJ, 489, 63\\
Guiderdoni, B.  \& Rocca-Volmerange, B.  1987, A\&A, 186, 1\\
Han, \etal 1998 ApJ, in press\\
Heckman, T. M., \etal 1998 preprint\\
Ichikawa, S. -I., Wakamatsu, K. -I., \& Okamura, S. 1986 ApJS 60, 475\\
Impey, C. \& Bothun, G. 1997, ARA\&A 35, 267\\
Impey, C. Sprayberry, D.. Irwin, M. J., \& Bothun, G. D. 1996, ApJS, 105, 209\\
Impey, C. \& Bothun, G. 1989, ApJ 341, 89\\
Kennicutt,  R. 1989, ApJ 344, 685 \\
Kennicutt, R., \etal 1987, AJ, 93, 1011\\
Knezeck, P., 1993 Ph.D. Thesis University of Pennsylvania\\
Larson, R., \& Tinsley, B. 1978, AJ, 219, 46\\
Linder, Suzanne 1998 1998 ApJ 495, 637\\
McGaugh, S., Bothun, G., \& Schombert, J. 1995, AJ 110, 573\\
McGaugh, S. 1994, ApJ 426, 135\\
McGaugh, S., \& Bothun, G. 1994, AJ, 107, 530\\
McGaugh, S., \& de Blok, W. 1993, AJ 106, 548\\
McGaugh, S. 1992, Ph.D. thesis, University of Michigan, Ann Arbor\\
Mengel, J., \etal 1979, ApJS, 40, 733\\
Mihos, C., \& Bothun, G. 1998 ApJ, in press\\
Mihos, C., de Block, W., \& McGaugh, S. 1997, ApJ, 477, L79\\
Mihos, C. \& Hernquist, L.  1996, ApJ, 464, 641\\
Mihos, C. \& Hernquist, L.  1994, ApJ, 425, L13\\
Mihos, C., Bothun, G., \& Richstone, D.  1993, ApJ, 418, 82\\
Mihos, C., Richstone, D., \& Bothun, G. 1991, ApJ, 377, 72\\
Mo, H. J., McGaugh, Stacy S., Bothun, Gregory D. 1994 MNRAS 267, 129\\
Oey, M.S. \& Clarke, C.J. 1998 AJ, 115, 1543\\
Oey, M.S. \& Kennicutt, R.C. Jr. 1997 MNRAS, 289, 570\\
O'Neil, K., 1998 in preparation\\
O'Neil, K., Carollo, \& Bothun, G.D. 1998, in preparation\\
O'Neil, K., Bothun, G., Impey, C., McGaugh, S. 1998a, AJ, 119 \\
O'Neil, K., Bothun, G., \& Impey, C. 1998b, AJ, submitted\\
O'Neil, K. 1997, Ph.D. thesis, University of Oregon, Eugene\\
O'Neil, K., Bothun, G., Schombert, J., Cornell, M., \& Impey, C. 1997a, AJ 114, 2448\\
O'Neil, K., Bothun, G., \& Cornell, M. 1997b, AJ 113, 1212\\
Pickering, T., Impey, C., Van Gorkom, J, Bothun, G. 1997, AJ, 114, 1858\\
Schaller, G., \etal 1992, A \& AS, 96, 269\\
Schombert, James M., Bothun, Gregory D., Schneider, Stephen E., \& McGaugh, Stacy S.
1992 AJ, 103, 1107\\
Salzer, J. 1998, preprint\\
Schombert J., \etal 1990, AJ, 100, 1523\\
Skillman, E. D., 1998, preprint\\
Skillman, E. D., Bothun, G. D., Murray, M. A., \& Warmels, R. H. 1987 A\&A, 185, 61\\
Smith, Denise A., \etal 1996 ApJ 473L, 21\\
Spitzak, \& Schneider 1998, preprint\\
Sprayberry, D., Impey, C. D., Irwin, M. J., \& Bothun, G. D. 1997a ApJ 482, 104\\
Sprayberry, D., Bernstein, G.M., Impey, C. D., \& Bothun, G. D. 1997b ApJ 438, 72\\
Sprayberry, D., Impey, C. D., Bothun, G. D., Irwin, M. J. 1995 AJ 109, 558\\
Sprayberry, D., Impey, C. D., Bothun, G. D., \& Irwin, M. J. 1995 AJ, 109, 558\\
Sweigart, A. \& Gross, P. 1978, ApJS, 36, 405\\
Van Der Hulst, J. M., Skillman, E. D., Smith, T. R., Bothun, G. D., McGaugh, S. S., \&
de Blok, W. J. G.  1993 AJ 106, 548\\
Van Zee, Liese, Haynes, Martha P., Salzer, John J., \& Broeils, Adrick H. 1997 AJ, 113, 1618\\
V\'{a}zquez, E. \& Scalo, J.  1989, ApJ, 343, 644\\
Webster, R., \etal  1998, preprint\\
\clearpage

Figure~\ref{fig:finalgas}.  The before and after gas density for various models.
Figure~\ref{fig:finalgas}(a) shows model a2 before (solid line) and after (dashed line)
the starburst as well as model l2 (dash-dotted line before and dotted line after).
Figure~\ref{fig:finalgas}(b) shows models a3 (solid line before and dashed line after)
and l3 (dash-dotted line before and dotted line after).  Finally, Figure~\ref{fig:finalgas}(c)
shows models i3 (solid line before and dashed line after)
and l4 (dash-dotted line before and dotted line after).

Figure~\ref{fig:concent}. The starburst distribution  
for four of the models -- c2, c5, f1, and f2.  The shading is proportional to the
total mass involved in the starburst.  Thus the darkest regions on the plot typically
lie at the core of the starburst while the white regions were unaffected by the starburst
(no gas was converted to stars).
Models c2 and c5 show the effects of
moving the burst radius from 0.10R$_T$ to 0.80R$_T$, while models f1 and f2 show
the effects of altering the concentration parameter c from 1.0 to 5.0.

Figure~\ref{fig:stars}. The distribution of stellar masses for four different models -
model e2 (solid line), model e1 (dashed line), model e3 (dash-dotted line), 
and model j5 (dotted line).  Model e2 (solid line) is representative of the majority
of the models.

Figure~\ref{fig:fig1}. The color and surface brightness plots for each of the models.
The top plot is the change in total U$-$B color, with time, for each of the models.
The bottom two plots show the surface brightness profile and U$-$B color profile of
each model 2 Myr after the burst.

Figure~\ref{fig:gal}.
Examples of surface brightness and color profiles of two LSB galaxies -- P2-4 (\muo = 25.14
\mss, Figure~\ref{fig:gal}(a) \& (b)), and P3-3 (\muo = 23.22 \mss, Figure~\ref{fig:gal}(c) \& (d))
(from O'Neil, \etal, 1997a, 1997b).

\clearpage
Table~\ref{tab:paramch}. Results for the one cell models  
discussed in section 3.1 (Table~\ref{tab:model}).
The numbers given are the change in total color and magnitude between t=0 (before the starburst)
and t=2 Myr.  

Table~\ref{tab:model}.  The model parameters for all models discussed in this paper.

Table~\ref{tab:param}. Results for the models discussed in this paper (Table~\ref{tab:model}).
The numbers given are the change in total color and magnitude between t=0 (before the starburst)
and t=2 Myr.

Table~\ref{tab:modell}.  Comparison of the results for models a1-a3 \& i3 with
models l1-l4.  
The numbers given are the change in total color and magnitude between t=0 (before the starburst)
and t=2 Myr.
\clearpage
\begin{table}
\caption{\label{tab:paramch}}
\end{table}
\begin{table}
\caption{\label{tab:model}}
\end{table}
\begin{table}
\caption{\label{tab:param}}
\end{table}
\begin{table}
\caption{\label{tab:modell}}
\end{table}

\clearpage
\clearpage

\begin{figure}[ht]
%\centerline{finalgas.ps}
\caption{\label{fig:finalgas}}
\end{figure}

\begin{figure}[ht]
%\centerline{concent.ps}
\caption{\label{fig:concent}}
\end{figure}

\begin{figure}[ht]
%\centerline{stars.ps}
\caption{\label{fig:stars}}
\end{figure}

\begin{figure}[ht]
%\centerline{
%\epsffile{fig1.ps}}
\caption{\label{fig:fig1}}
\end{figure}

\begin{figure}[ht]
%\centerline{
%\epsfysize=8.5in
%\epsffile{bp2-4.ps}}
\caption{a}
\label{fig:gal}
\end{figure}
\addtocounter{figure}{-1}
\begin{figure}[ht]
%\centerline{
%\epsfysize=8.5in
%\epsffile{bvp2-4.ps}}
\caption{b}
\end{figure}
\addtocounter{figure}{-1}
\begin{figure}[ht]
%\centerline{
%\epsfysize=8.5in
%\epsffile{bp3-3.ps}}
\caption{c}
\end{figure}
\addtocounter{figure}{-1}
\begin{figure}[ht]
%\centerline{
%\epsfysize=8.5in
%\epsffile{ubp3-3.ps}}
\caption{d}
\end{figure}

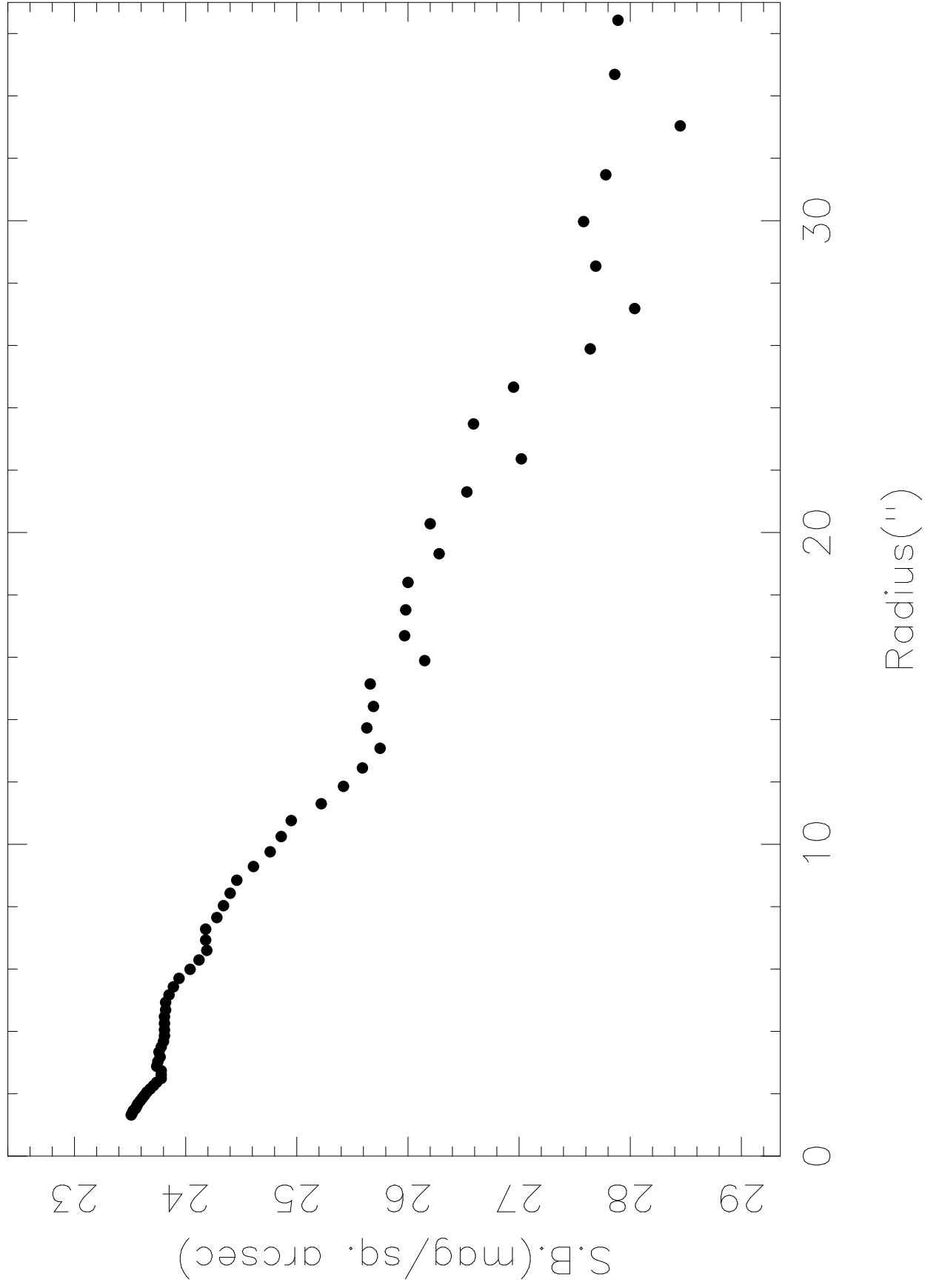
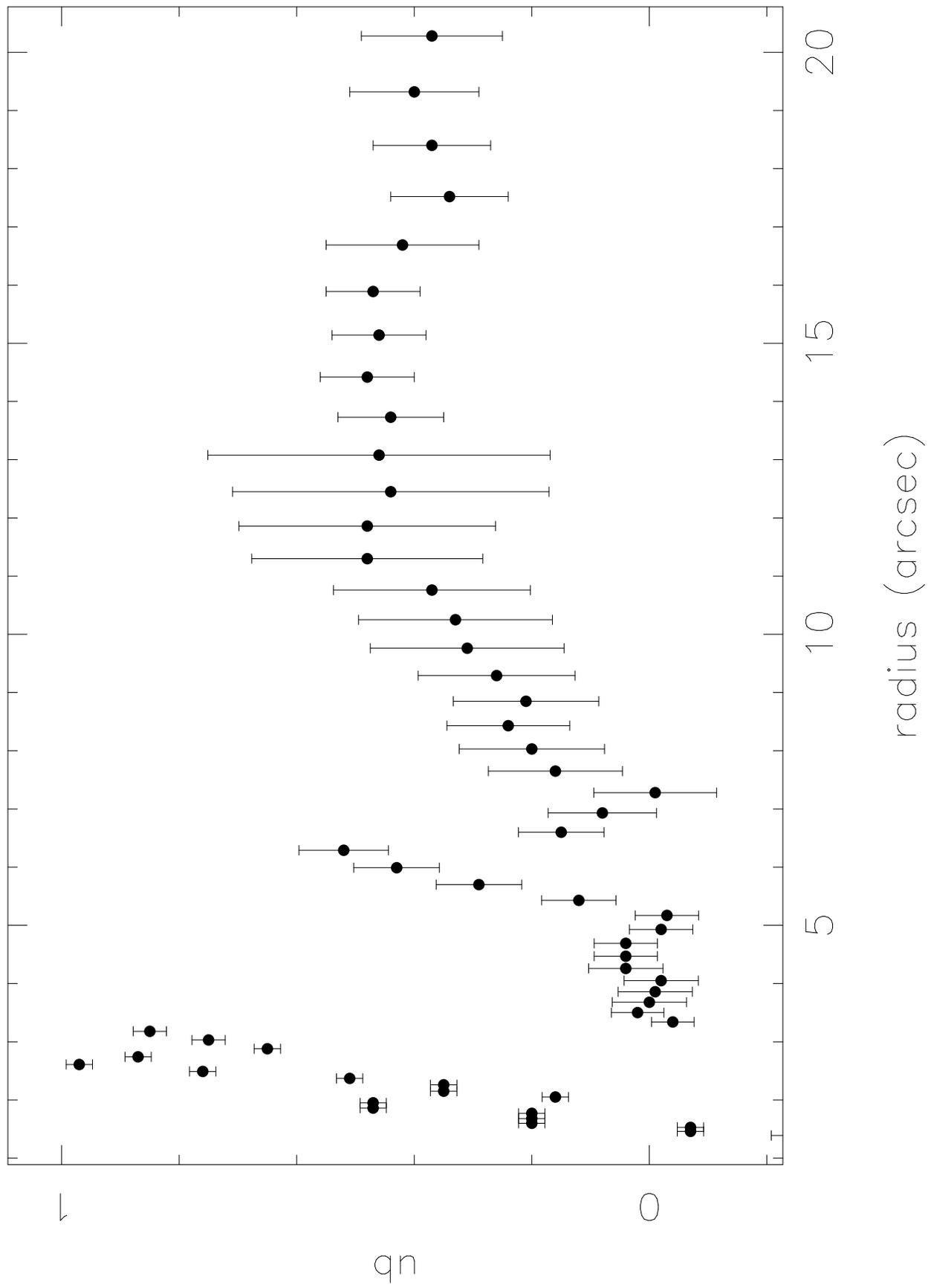
\end{document}